\documentclass[12pt]{iopart}
\usepackage{graphicx}
\usepackage{color}

\begin{document}

\title[]{Structure of global buyer-supplier networks and its implications for conflict minerals regulations}

\author{Takayuki Mizuno$^{1,2,3,6}$\footnote{Corresponding 
author: mizuno@nii.ac.jp}, Takaaki Ohnishi$^{4,6}$, and Tsutomu Watanabe$^{5,6}$}

\address{$^1$National Institute of Informatics, Hitotsubashi, Chiyoda-ku, Tokyo, 101-8430, JAPAN}
\address{$^2$Department of Informatics, SOKENDAI (The Graduate University for Advanced Studies), Hitotsubashi, Chiyoda-ku, Tokyo, 101-8430, JAPAN}
\address{$^3$PRESTO, Japan Science and Technology Agency, Hitotsubashi, Chiyoda-ku, Tokyo, 101-8430, JAPAN}
\address{$^4$Graduate School of Information Science and Technology, University of Tokyo, Hongo, Bunkyo-ku, Tokyo, 113-8656, JAPAN}
\address{$^5$Graduate School of Economics, University of Tokyo, Hongo, Bunkyo-ku, Tokyo, 113-0033, JAPAN}
\address{$^6$The Canon Institute for Global Studies, Marunouchi, Chiyoda-ku, Tokyo, 100-6511, JAPAN}

\ead{mizuno@nii.ac.jp}

\begin{abstract}
We investigate the structure of global inter-firm linkages using a dataset that contains information on business partners for about $400,000$ firms worldwide, including all the firms listed on the major stock exchanges. Among the firms, we examine three networks, which are based on customer-supplier, licensee-licensor, and strategic alliance relationships. First, we show that these networks all have scale-free topology and that the degree distribution for each follows a power law with an exponent of $1.5$. The shortest path length is around six for all three networks. Second, we show through community structure analysis that the firms comprise a community with those firms that belong to the same industry but different home countries, indicating the globalization of firms' production activities. Finally, we discuss what such production globalization implies for the proliferation of conflict minerals (i.e., minerals extracted from conflict zones and sold to firms in other countries to perpetuate fighting) through global buyer-supplier linkages. We show that a limited number of firms belonging to some specific industries and countries plays an important role in the global proliferation of conflict minerals. Our numerical simulation shows that regulations on the purchases of conflict minerals by those firms would substantially reduce their worldwide use.
\end{abstract}

\vspace{2pc}

\maketitle

\section{Introduction}
Many complex physical systems can be modeled and better understood as complex networks \cite{Watts, Albert, Helbing}. Recent studies show that economic systems can also be regarded as complex networks in which economic agents, like consumers, firms, and governments, are closely connected \cite{Jackson, Goyal}. To understand the interaction among economic agents, we must uncover the structure of economic networks. 

Our focus in this paper is on inter-firm networks. Over the last several years, a pair of critical incidences revealed the global interconnection of firms: the financial turmoil triggered by the subprime mortgage crisis in the United States and the disruption of worldwide supply chains caused by the 2011 Fukushima earthquake and tsunami in Japan. These incidences sparked interest in inter-firm networks by network scientists, economists, and sociologists. Several aspects of inter-firm relationships were previously studied, including the firm and bank relationship \cite{Souma}, the customer-supplier relationship \cite{Saito, Fujiwara, Luo, Atalay, Takayasu}, and the ownership relationship (i.e., which firm owns which firm) \cite{Glattfelder, Kogut}). The degree distribution follows a power law, and the shortest path-lengths for any pair of firms are around five \cite{Takayasu, Ohnishi, Mizuno}. Newman's community analysis \cite{Newman_03, Newman_04} also shows that firms tend to transact with each other when they belong to the same industry and/or are located in the same state or prefecture \cite{Iino}. The structure of the inter-firm networks investigated was found to be stable \cite{Mizuno}. 

One limitation of previous studies is that they mainly addressed inter-firm relationships within a country, so they have little to say about the linkages of firms in different countries. Important exceptions include studies on global ownership structure \cite{Vitali}, and global R\&D partnerships \cite{Bojanowski}. However, no work has yet investigated the structure of global customer-supplier linkages, which is the key to understanding the nature of the above incidences (i.e., the global financial crisis and the disruption of global supply chains by natural disasters). Given this background, we investigate the global aspect of inter-firm linkages using a unique dataset that contains information on business relationships (both domestic and international) for about $400,000$ firms worldwide. This is the first contribution of this paper. Note that if one aggregates transactions (purchases and sales) at the firm level, one obtains transactions at the country level, called international trade by economists. The network structure of international trade has recently been studied by economists and physicists \cite{Garlaschelli_05, Garlaschelli_04, Giovanni, Piccardi, He, Barigozzi}. Our investigation of international transactions at the firm level is closely related to those studies, but our firm level data reveal that firms are more likely to transact with other firms in the same industry rather than other firms in the same home country, which is quite different from what is assumed in studies on international trade. We will expand this point below. 

After examining the structure of global inter-firm networks, we discuss the implications of global linkages at the firm level for the proliferation of "conflict minerals" through global buyer-supplier linkages. This is the second contribution of this paper. Conflict minerals are natural minerals (gold, tin, tungsten, etc.) that are extracted from conflict zones and sold to perpetuate fighting. The most prominent example is the natural minerals extracted in the Democratic Republic of the Congo (DRC) by armed groups and funneled through a variety of intermediaries before being purchased by multinational electronics firms in industrial countries. There is wide discussion on how to mitigate the worldwide spread of conflict minerals \cite{Ross, Billon_13, Billon_01}. For example, the U.S. government passed a law in July 2010 that requires listed firms to audit their supply chains and report the use in their products of conflict minerals from the DRC or adjoining countries \cite{Section_1502}. In this paper, we conduct simulation analysis using our firm level data on global buyer-supplier linkages to evaluate the effectiveness of various measures to mitigate the worldwide propagation of conflict minerals through global supply chains.

The rest of the paper is organized as follows. Section 2 provides a detailed description of our dataset, and section 3 looks at the basic features of inter-firm networks, including degree distributions. Section 4 examines how closely firms are interconnected by measuring the shortest path lengths between them. Section 5 conducts community analysis employing a network's modularity defined by Newman. Section 6 conducts simulation analysis to investigate the proliferation of conflict minerals through global supply chains. Section 7 concludes the paper.

\section{Dataset}
For this research, we used a unique dataset compiled by S\&P Capital IQ, which is part of McGraw Hill Financial Inc. \cite{Capital_IQ}. The dataset covers $423,024$ major incorporated firms with information of business relationships in $217$ countries in $159$ industry sectors that were defined by the S\&P, including all the listed firms in the world. These firms have such attributions as location, industry sector, and annual total revenue. The dataset has lists of core partners (i.e., customers, suppliers, licensors, licensees, and strategic alliances) in recent period (i.e., 2013 and/or 2014) and in some years over the entire sample period (almost all relationships date from 2000) for a firm with IDs. For example, the numbers of core customers and suppliers for IBM are $446$ and $184$ firms in the recent period and $1565$ and $755$ firms in some years over the entire sample period. The following relationship exists between customers and suppliers. When firm $j$ is a supplier to firm $i$, firm $i$ is a customer of firm $j$. Licensees and licensors have such an opposite relationship.

\section{Scale-free global inter-firm networks}
We first show the common scale-free properties of three global inter-firm networks, which are based on customer-supplier, licensee-licensor, and strategic alliance relationships. Figure 1 describes a global customer-supplier network. Many international linkages through U.S., European, and Japanese firms are observed. Licensee-licensor and strategic alliance networks have identical characteristics. We investigate the cumulative distribution functions (CDFs) of links across firms for the following linkages: customer, supplier, licensee, licensor, and strategic alliance. In figure 2, the horizontal axis is the number of links, and the vertical axis represents the cumulative densities. The CDFs of each kind of linkage have tails that follow a power law function. The power law exponents are about $1.5$ as follows:

\begin{equation}
{\rm Supplier:}		P_> (N_s) \propto N_s^{-1.5},
\end{equation}
\begin{equation}
{\rm Customer:}		P_> (N_c) \propto N_c^{-1.5},
\end{equation}
\begin{equation}
{\rm Licensee:}		P_> (N_{le}) \propto N_{le}^{-1.5},
\end{equation}
\begin{equation}
{\rm Licensor:}		P_> (N_{lo}) \propto N_{lo}^{-1.5},
\end{equation}
\begin{equation}
{\rm Strategic \ alliance:}  	P_> (N_{sa}) \propto N_{sa}^{-1.5},
\end{equation}

\noindent
where $N$ is the number of links for each type of linkage. The $1.5$ exponent is observed on the distribution both in the recent period and in some years over the entire sample period. Networks with such power laws are called scale-free networks. 

\section{Six degrees of separation between major incorporated firms}
We measure the shortest path lengths (SPLs) for each pair of firms $i$ and $j$ in a non-directed customer-supplier network over the entire sample period. The number of firms on it is $345,909$, so there are about $60$ billion pairs. Although the density for the linkages in this network is very low, only $1.1 \times 10^{-5}$, the number of firms in its maximum connected component is $318,080$, which is about $92.0\%$ of all the firms. Figure 3 shows the SPL distribution for the connected pairs. The mode of distribution is five path lengths, and about $80\%$ of the pairs are connected by six path lengths or fewer. We also investigate the mode for the directed customer-supplier network to consider intermediate products or money flow. This mode is also short, only seven path lengths. We also measure the SPLs in non-directed licensee-licensor, directed licensee-licensor, and non-directed strategic alliance networks. The number of firms, the network density for the linkages, the size of the maximum connected component, and the mode of the SPLs are shown in table 1. The mode in these networks is also short: around six path lengths.

The SPL distribution of a domestic non-directed customer-supplier network was previously reported \cite{Takayasu, Ohnishi, Mizuno}. In the Japanese case, the mode of SPL distribution is five path lengths. International linkages for firms were not taken in these previous studies; the actual SPL for the Japanese firm's pairs is shorter than five path lengths. We compare SPLs for firm's pairs in the same country in a global customer-supplier network and a domestic customer-supplier network that has only domestic linkages. Figure 4 shows the probability that SPLs are extended by being limited to domestic linkages in the United States or Japan. In the Japanese case of ${\rm SPL}\ge5$ and the U.S. case of ${\rm SPL}\ge7$ in a global customer-supplier network, the probability is over $0.5$. This result suggests that the effect from foreign countries cannot be ignored, because international relationships are indispensable for handling incidents in supply chains.

\section{Firms' community structure}
Global inter-firm networks are built by multiple communities. Detecting the communities in networks means the appearance of dense connected groups of vertices and sparse connections among groups. We adopt modularity as a quality function of communities introduced by Newman and detect them by a fast greedy algorithm of modularity maximization that is one effective approach to identify communities \cite{Clauset}. If network $V$ is divided into $L$ subsets $\{V_1,V_2,\cdots,V_L\}$ which do not overlap and are not empty, modularity $Q$ is defined as

\begin{equation}
Q=\sum_{i=1}^L (e_{ii}-a_{i}^2)=\frac{1}{2M}\sum_{l \in V_i}\sum_{m \in V_i}A_{lm}-\left(\frac{1}{2M}\sum_{l \in V_i}\sum_{m \in V_i}A_{lm}\right)^2,
\end{equation}

\noindent
where $A_{lm}$ is an element of adjacent matrix which indicates the number of links between nodes $l$ and $m$. $e_{ii}$ and $a_i$ are the link densities within subset $V_i$ and the number of links that connect in to subsets $V_i$, respectively. When the subsets $\{V_1,V_2,\cdots,V_L\}$ are selected randomly, $e_{ii}$ is canceled out by $a_i^2$, which gives the expectation value of the network density for the linkages in subset $V_i$. Using the modularity we can compare the actual network density for linkages in a subset with its expectation value. When $Q\simeq 0$, the network has no statistically significant communities, unlike randomly connected networks. On the other hand, $Q\simeq 1$ corresponds to a network which is almost perfectly partitioned into modules.

The maximum modularities of non-directed customer-supplier, non-directed licensor-licensee, and strategic alliance networks over the entire sample period are $0.64$, $0.75$, and $0.74$, respectively. Such sufficiently large modularity means that significant communities exist in the networksDWe characterize each community by checking the majority attributes (e.g., country and industry sector) of the firms in the community. Because firm attribution bias can be found in each network, we compare the fraction of the firms' attribution in each community with the fraction in all communities by $R_{i,l}$, which is defined as

\begin{equation}
R_{i,l}=\frac{ {\rm fraction \: of \: attribution}\ i {\rm \: in \: community}\ l {\rm \: in \: network} }{ {\rm fraction \: of \: attribution}\ i {\rm \: in \: all \: communities \: in \: network} }.
\end{equation}

\noindent
The $p$-values for the fraction of the firms' attribution in each community are calculated using the null hypothesis that the firms' attribution is independent of community.

First, we investigate the communities in the licensee-licensor network with the highest modularity over the entire sample period. The network has $4,493$ communities. However, the top four account for over $50\%$ of all firms: $22.9\%$, $18.3\%$, $7.7\%$, and $3.8\%$. Table 2 shows the fraction of firms' attributions with a $p$-value $< 0.01$ among the top four communities. We focus on the remarkable attributions with $R_{i,l}\ge 3$ in each attribution to identify the community characteristics. The largest community is mainly comprised of movies, entertainment, semiconductors, and broadcasting firms. Many Taiwan firms appertain to this community. The major industry in Taiwan is semiconductors. Therefore, the largest community expresses sectors for broadcasting technology. As shown in table 2, the 2nd, 3rd, and 4th largest communities show the ICT sectors in health care, the apparel sectors, and the chemical industry sectors, respectively. The licensee-licensor relationships between major incorporated firms tend to be confined to similar industrial sectors over the boundaries between countries.

We next focus on over the entire sample period a non-directed strategic alliance network which has $1,995$ communities. The top four account for $18.1\%$, $17.2\%$, $10.8\%$, and $9.5\%$ of all the firms. In each community, the firms belong to the similar industrial sectors but different home countries. The largest community mainly includes firms in ICT sectors (i.e., IT consulting and other services, communications equipment, system software) as shown in table 3. The 2nd, 3rd, and 4th largest communities express heavy industry, bank and resort development, and medicine sectors, respectively.

We also investigate firms' attribution in each community in a non-directed customer-supplier network over the entire sample period. The network has $3,463$ communities. The top four account for $20.9\%$, $20.7\%$, $12.0\%$, and $5.9\%$ of all the firms. The 2nd and 3rd largest communities show industry sectors, such as aerospace/defense and health care (table 4). On the other hand, the 4th largest community shows transactions in the ASEAN free trade area because Southeast Asian firms tend to densely connect to firms in the same area. Since various industries are included in the largest community (table 4), we further divide the largest one into discrete sub-communities. The major sub-communities express some industry sectors. The 1st, 2nd, and 3rd largest sub-communities show the broadcasting technology sectors, department stores (i.e., apparel and restaurant sector), and the electronic equipment sectors, respectively (table 5).

As cited above, major incorporated firms tend to have worldwide connections. We investigate the relationship between firm size and geographical distance to business partner. Here, firm size is measured by the total 2013 revenue. As shown in figure 5, the mean of the geographical distance in a customer-supplier network is about $3,400$ km, which is shorter than in other inter-firm networks (i.e., $3,700$ km for strategic alliance relationships and $4,300$ km for licensee-licensor relationships). Because firms choose suppliers and customers by taking into consideration transport costs and product price, the mean of the geographical distance of the customer-supplier network is short. As represented by the 4th largest community in the customer-supplier network, a large community that expresses a region is only observed in this network. In inter-firm networks, the geographical distance of a large firm whose annual total revenue exceeds $10^3$ million dollars tends to be long; large firms are affected by the economic conditions in distant unexpected countries.

\section{Simulation for conflict minerals proliferation}
We follow the recent literature on supply chains \cite{Watanabe, Foerster} and introduce the following simple diffusion model that resembles PageRank:

\begin{equation}
\fl
\left(
\begin{array}{c}
g_{1}(t+1)\\
g_{2}(t+1)\\
\vdots \\
g_{N}(t+1)
\end{array}
\right)
=
\left(
\begin{array}{c@{}c@{}c@{}c}
1-q_{1} & 0 & \cdots & 0 \\
0 & 1-q_{2} & \cdots & 0 \\
\vdots & \vdots & \ddots & \vdots \\
0 & 0 & \cdots & 1-q_{N}
\end{array}
\right)
\left(
\begin{array}{c@{}c@{}c@{}c}
0 & a_{12} & \cdots & a_{1N} \\
a_{21} & 0 & \cdots & a_{2N} \\
\vdots & \vdots & \ddots & \vdots \\
a_{N1} & a_{N2} & \cdots & 0
\end{array}
\right)
\left(
\begin{array}{c}
g_{1}(t) \\
g_{2}(t) \\
\vdots \\
g_{N}(t) 
\end{array}
\right)
+
\left(
\begin{array}{c}
\epsilon_{1}(t) \\
\epsilon_{2}(t) \\
\vdots \\
\epsilon_{N}(t) 
\end{array}
\right)
\end{equation}

\noindent
We explain this general model using an example for conflict minerals. $g_i(t)$ is the amount of conflict minerals possessed by firm $i$ at time $t$, $\epsilon_i(t)$ is the amount of conflict minerals that are extracted by firm $i$ at time $t$ that doesn't stem from customer-supplier chains, $q_i$ expresses the rate at which they are consumed as a port of the final consumption products in firm $i$, and $a_{ij}$ is an element in an input-output matrix. Typical element $a_{ij}$ equals $1/\hat{N}_j^C$ if firm $i$ is a customer of firm $j$ and zero otherwise. In the standard notation adopted in input-output analysis, $a_{ij}$ represents the share of output $j$ (i.e., commodity produced by firm $j$) in the total intermediate output use of firm $i$. We have information on whether firm $i$ purchases something from firm $j$, but no information on the amount of output $j$ purchased by firm $i$; this is the thickness of each customer link. Since we assume that the customer links of firm $j$ have identical thickness, $a_{ij}=1/\hat{N}_j^C$ if firm $i$ is a customer of firm $j$ and zero otherwise. From PageRank theory, it is trivial that $g_i(t)$ converges on $g_i(t-1)$ at time $t=\infty$ when all $q_i$ satisfy the inequality, $0<q_i\le 1$.

We simulate the diffusion of the conflict minerals that are mined in the Democratic Republic of the Congo (DRC) and all of its nine neighbors: Angola, Burundi, Central African Republic, Republic of the Congo, Rwanda, South Sudan, Tanzania, Uganda, and Zambia. We simply assume that $\epsilon_{i}(t)$ is time independent and set the total 2013 revenue of each firm in the ``metals and mining'' upper-sector (which includes aluminum, diversified metals and mining, gold, precious metals and minerals, silver, and steel sectors) in these countries to $\epsilon_{i}(t)$. Here, for  $\epsilon_{i}(t)$ for some firms whose total revenue is not recorded in this dataset, we substitute the mean of the total revenue of the firm whose total revenue is recorded in this upper-sector and these countries. Except for these firms,  $\epsilon_{i}(t)=0$. All initial values are $g_{i}(0)=0$.

We ran the model until $g_{i}(t)$ practically converges on $g_{i}(t-1)$ on the customer-supplier network without banking sectors over the entire sample period. Figure 6 shows a simulated amount of the conflict minerals per firm in each country when all $q_{i}=0.3$. The conflict minerals drift down to the firms even in most developed countries. In the G8 countries the top ten industry sectors in which many conflict minerals hide products are shown in table 6. Conflict minerals are found in the ``electrical components and equipment'' sector in G8. The ``metals and mining'' upper-sector and the ``trading companies and distributors'' sector have only $3.4\%$ of all the firms in the G8 and account for $94.6\%$ of the total conflict minerals in G8.

We numerically demonstrate the simplified regulation that all firms in the ``metals and mining'' upper-sector and the ``trading companies and distributors'' sector in G8 must not distribute conflict minerals to their customers; the $q_{i}$s of these firms are one and $q_{i}=0.3$ otherwise. As shown in table 6, the amount of conflict minerals is reduced in each sector. The conflict minerals in the ``electrical components and equipment'' and ``alternative carriers'' sectors fell by over $97\%$, where they substantially improved their conflict minerals issues.

We numerically show that the amount of conflict minerals would decrease effectively outside the G8 by a regulation on the purchases of conflict minerals directed at $3\%$ of all the firms in G8. We selected the $3\%$ firms in each of the following conditions and dammed the conflict mineral flow; their $q_{i}$s are one and $q_{i}=0.3$ otherwise.

\begin{description}
\item[Condition 1]Firms are selected in descending order of the number of supplier links in G8.
\item[Condition 2]Only listed firms are selected in descending order of the number of supplier links in G8.
\item[Condition 3]Firms are selected in descending order of the number of supplier links in the ``metals and mining'' upper-sector and ``trading companies and distributors'' sectors in G8.
\end{description}

Table 7 shows the numerical simulation results of the regulation with each condition. In condition 1, the selected $3\%$ firms decreased distribution of all conflict minerals by $35.0\%$ in G8. However, the amount of conflict minerals hardly changed in the other firms. A U.S. federal law for conflict minerals among the listed firms was passed on July 21, 2010 \cite{Section_1502}. Condition 2 expresses the situation where this law is applied to all the listed firms in G8. Although the trend improved slightly from $35.0\%$ to $43.2\%$ in G8, the amount of conflict minerals also hardly decreased outside of the selected firms. On the other hand, we confirmed a dramatic reduction in condition 3. The selected $3\%$ of firms decreased distribution of all conflict minerals by $97.3\%$ in G8 and created a block that intercepted about $12.0\%$ of all the conflict minerals outside of G8, DRC, and DRC's neighbors.

\section{Conclusion}
We investigated the structure of global inter-firm relationships using a unique dataset that contains the information of business relationships for $423,024$ major incorporated firms and focused on three different networks: a customer-supplier network through which products and services flow; a licensee-licensor network through which technical information and know-how flow; and a strategic alliance network through which both flow mutually.

These networks have common scale-free properties. The degree distributions follow a power law with an exponent of $1.5$. The shortest path length for each pair of firms is around six for all three networks. We showed through community structure analysis that the firms comprise a community with those firms that belong to the same industry but different home countries, indicating the globalization of firms' production activities.

We measured the mean of the geographical distance between the firms and their business partners. It was $3,400$ km between business partners for customer-supplier relationships, which is shorter than for the two other relationships. This result suggests that technical information and know-how without high transport costs have the potential to be diffused rapidly worldwide. We also confirmed that the geographical distance between business partners for large firms tends to be long.

Finally, by utilizing a simple diffusion model and empirical results where firms comprise a community with those firms that belong to the same industry but different home countries, we showed numerically that regulations on the purchases conflict minerals by limited number of G8 firms belonging to some specific industries would substantially reduce their worldwide use. When these firms refuse to buy conflict minerals from their suppliers, the supply chains of many intermediaries which are positioned upstream suffer. Future work will accurately estimate each intermediary's amount of damage and the model's parameters by comprehensively collecting the data of global inter-firm relationships. Part of the money which was spent in firms flows into conflict minerals through multiple inter-firm networks. Future work will also expand our model by connecting a customer-supplier network with licensee-licensor and strategic alliance networks. This expanded model might help make more effective policies for conflict minerals. Recently, the global diffusion of weaponry, technical know-how, conflict oil, and natural gas through lawful trades is also attracting attention. Our results might resolve such issues and contribute to global peace.

\ack
We express our deep appreciation to Professors Hiroshi Iyetomi and Yuichi Ikeda who provided insightful comments and suggestions. This work was partially supported by the Ishii Memorial Securities Research Promotion Foundation and JSPS KAKENHI Grant Number 24710156.

\clearpage
\section*{Tables}

\begin{table}[!ht]
\caption{\label{tabl_1}Number of firms, network density for linkages, maximum connected component size, and mode of shortest path lengths in customer-supplier (CS), licensee-licensor (LL), and strategic alliance (SA) networks in recent period (i.e., 2013 and/or 2014) and over entire sample period.}
\footnotesize\rm
\begin{tabular*}{\textwidth}{crllllll}
\br
&&\centre{2}{Density}&&\centre{2}{Mode}\\
\ns
&&\crule{2}&\% of max.&\crule{2}\\
&\# of firms&Non-directed&Directed& conected component&Non-directed&Directed\\
\mr
CS net.$^{\rm a}$&$345,909$&$1.1\times10^{-5}$&$5.7\times10^{-6}$&$92.0\%$&$5$&$7$\\
CS net.$^{\rm b}$&$123,052$&$2.5\times10^{-5}$&$1.3\times10^{-5}$&$85.4\%$&$6$&$8$\\
LL net.$^{\rm a}$&$36,264$&$6.5\times10^{-5}$&$3.6\times10^{-5}$&$60.4\%$&$6$&$8$\\
LL net.$^{\rm b}$&$12,646$&$1.6\times10^{-4}$&$8.5\times10^{-5}$&$54.3\%$&$8$&$8$\\
SA net.$^{\rm a}$&$124,444$&$2.3\times10^{-5}$&--&$77.8\%$&$6$&--&\\
SA net.$^{\rm b}$&$47,877$&$4.9\times10^{-5}$&--&$64.8\%$&$6$&--&\\
\br
\end{tabular*}
$^{\rm a}$ Entire sample period.
$^{\rm b}$ Recent period.
\end{table}

\begin{table}[!ht]
\caption{\label{tabl_2}Top 5 fractions of firms' attributes with $p$-value $< 0.01$ in major communities in non-directed licensee-licensor network. $R$, which expresses ratio between actual fraction and the fraction obtained by random selection, is defined by equation (7). Bold font indicates remarkable attributes with $R\ge3$.}
\footnotesize\rm
\begin{tabular*}{\textwidth}{cll}
\br
Rank&Country	(fraction ($> 0.01$), $R$)&Industry sector (fraction ($> 0.01$), $R$)\\
\mr
1	&United Kingdom ($0.067$, $1.18$)		&{\bf Movies and entertainment} ($0.105$, $3.68$)\\
	&Japan ($0.043$, $1.50$)			&{\bf Semiconductors} ($0.083$, $4.16$)\\
	&{\bf Taiwan} ($0.033$, $3.34$)		&{\bf Broadcasting} ($0.076$, $4.07$)\\
	&France ($0.026$, $1.31$)		         &Internet software and services ($0.061$, $1.55$)\\
	&South Korea ($0.020$, $1.72$)		&Application software ($0.056$, $1.19$)\\
\mr
2	&United States ($0.551$, $1.13$)		&{\bf Internet software and services} ($0.346$, $6.03$)\\
	&Japan ($0.047$, $1.65$)			&{\bf Application software} ($0.296$, $5.59$)\\
	&Germany ($0.040$, $1.44$)		&Asset Management and custody banks ($0.073$, $2.74$)\\
	&Switzerland ($0.029$, $2.50$)		&{\bf Health care technology} ($0.052$, $4.36$)\\
	&France ($0.028$, $1.37$)		         &Healthcare equipment ($0.029$, $2.34$)\\
\mr
3	&United States ($0.622$, $1.27$)		&{\bf Apparel, accessories and luxury goods} ($0.287$, $13.59$)\\
	&{\bf Italy} ($0.046$, $3.72$)			&{\bf Distributors} ($0.056$, $4.98$)\\
	&France ($0.032$, $1.60$)		         &{\bf Apparel retail} ($0.051$, $7.91$)\\
	&Thailand ($0.010$, $1.94$)		&{\bf Footwear} ($0.046$, $11.50$)\\
	&			&Packaged foods and meats ($0.032$, $1.55$)\\
\mr
4	&India ($0.046$, $1.51$)			&{\bf Commodity chemicals} ($0.128$, $9.50$)\\
	&Japan ($0.046$, $1.60$)			&{\bf Oil and gas refining and marketing} ($0.072$, $11.90$)\\
	&Netherlands ($0.027$, $2.43$)		&{\bf Fertilizers and agricultural chemicals} ($0.050$, $13.17$)\\
	&South Korea ($0.025$, $2.21$)		&Industrial machinery ($0.046$, $2.55$)\\
	&Israel ($0.023$, $2.23$)			&{\bf Construction and engineering} ($0.044$, $5.87$)\\
\br
\end{tabular*}
\end{table}

\begin{table}[!ht]
\caption{\label{tabl_3}Top 5 fractions of firms' attributes with $p$-value $< 0.01$ in major communities in non-directed strategic alliance network. Bold font expresses remarkable attributes with $R\ge3$.}
\footnotesize\rm
\begin{tabular*}{\textwidth}{cll}
\br
Rank&Country	(fraction, $R$)&Industry sector (fraction, $R$)\\
\mr
1	&United States ($0.445$, $1.22$)		&Application software ($0.150$, $2.83$)\\
	&United Kingdom ($0.068$, $1.15$)		&Internet software and services ($0.127$, $2.21$)\\
	&France ($0.036$, $1.48$)		         &{\bf IT consulting and other services} ($0.088$, $3.32$)\\
	&Taiwan ($0.018$, $2.25$)		&{\bf Communications equipment} ($0.062$, $3.66$)\\
	&South Korea ($0.014$, $12.0$)		&{\bf Systems software} ($0.059$, $3.79$)\\
\mr
2	&India ($0.072$, $2.06$)			&{\bf Oil and gas exploration and production} ($0.097$, $3.53$)\\
	&China ($0.071$, $1.30$)			&Construction and engineering ($0.075$, $2.18$)\\
	&United Kingdom ($0.067$, $1.13$)		&{\bf Aerospace and defense} ($0.042$, $3.56$)\\
	&Japan ($0.066$, $1.94$)			&Industrial machinery ($0.033$, $2.01$)\\
	&Australia ($0.052$, $1.56$)		&Electric utilities ($0.031$, $2.74$)\\
\mr
3	&China ($0.061$, $1.12$)			&{\bf Diversified banks} ($0.105$, $7.65$)\\
	&Japan ($0.060$, $1.75$)			&{\bf Regional banks} ($0.067$, $6.24$)\\
	&France ($0.036$, $1.46$)		         &Asset management and custody banks ($0.064$, $2.41$) \\
	&Indonesia ($0.028$, $2.79$)		&{\bf Airlines} ($0.056$, $8.90$)\\
	&Hong Kong ($0.026$, $1.96$)		&{\bf Hotel, resorts and cruise lines} ($0.048$, $5.27$)\\
\mr
4	&United States ($0.510$, $1.40$)		&{\bf Pharmaceuticals} ($0.205$, $9.35$)\\
	&Germany ($0.040$, $1.20$)		&{\bf Biotechnology} ($0.192$, $10.23$)\\
	&France ($0.029$, $1.19$)		         &{\bf Life sciences tools and services} ($0.075$, $9.11$)\\
	&Switzerland ($0.020$, $1.86$)		&{\bf Healthcare equipment} ($0.073$, $5.84$)\\
	&Sweden ($0.017$, $1.50$)		&{\bf Healthcare facilities} ($0.067$, $6.29$)\\
\br
\end{tabular*}
\end{table}

\begin{table}[!ht]
\caption{\label{tabl_4}Top 5 fractions of firms' attributes with $p$-value $< 0.01$ in major communities in non-directed customer-supplier networks. Bold font expresses remarkable attributes with $R\ge3$.}
\footnotesize\rm
\begin{tabular*}{\textwidth}{cll}
\br
Rank&Country	(fraction, $R$)&Industry sector (fraction, $R$)\\
\mr
1	&United States ($0.442$, $1.34$)		&Internet software and services ($0.081$, $2.26$)\\
&United Kingdom ($0.079$, $1.19$)		&Application software ($0.058$, $1.79$)\\
&Japan ($0.032$, $1.26$)			&Communications equipment ($0.032$, $2.33$)\\
&France ($0.026$, $1.32$)		         &IT consulting and other services ($0.030$, $1.55$)\\
&Taiwan ($0.023$, $2.27$)		&Regional banks ($0.030$, $2.18$)\\
\mr
2	&India ($0.103$, $2.58$)			&Construction and engineering ($0.064$, $1.63$)\\
&Australia ($0.035$, $1.24$)		&{\bf Aerospace and defense} ($0.048$, $3.82$)\\
&Japan ($0.033$, $1.32$)			&Oil and gas exploration and production ($0.041$, $2.82$)\\
&Germany ($0.031$, $1.06$)		&Industrial machinery ($0.039$, $1.73$)\\
&France ($0.024$, $1.19$)		         &Electric utilities ($0.038$, $2.66$)\\
\mr
3	&United States ($0.379$, $1.15$)		&{\bf Pharmaceuticals} ($0.086$, $5.34$)\\
&United Kingdom ($0.155$, $2.32$)		&{\bf Healthcare facilities} ($0.070$, $4.96$)\\
&{\bf Poland} ($0.069$, $5.52$)			&{\bf Healthcare equipment} ($0.056$, $4.71$)\\
&Sweden ($0.033$, $2.96$)		&{\bf Biotechnology} ($0.048$, $5.89$)\\
&Norway ($0.018$, $2.49$)		&{\bf Healthcare services} ($0.042$, $4.46$)\\
\mr
4	&{\bf Indonesia} ($0.340$, $14.91$)		&{\bf Property and casualty insurance} ($0.114$, $14.91$)\\
&{\bf Thailand} ($0.090$, $7.87$)		&Asset management and custody banks ($0.073$, $2.46$)\\
&{\bf Philippines} ($0.085$, $12.26$)		&{\bf Life and health insurance} ($0.058$, $9.13$)\\
&Singapore ($0.028$, $2.07$)		&Packaged foods and meats ($0.045$, $2.21$)\\
&Malaysia ($0.020$, $1.14$)		&{\bf Reinsurance} ($0.044$, $22.12$)\\
\br
\end{tabular*}
\end{table}

\begin{table}[!ht]
\caption{\label{tabl_5}Top 5 fractions of firm attributes with $p$-value $< 0.01$ in major sub-communities in largest community in customer-supplier network. Bold font expresses remarkable attributes with $R\ge3$.}
\footnotesize\rm
\begin{tabular*}{\textwidth}{cll}
\br
Rank&Country	(fraction ($> 0.01$), $R$)&Industry sector (fraction ($> 0.01$), $R$)\\
\mr
1	&United Kingdom ($0.078$, $1.38$)		&{\bf Movies and entertainment} ($0.135$, $4.73$)\\
&Israel ($0.017$, $1.66$)			&{\bf Broadcasting} ($0.128$, $6.83$)\\
	&			&Internet software and services ($0.093$, $2.33$)\\
	&			&Application software ($0.074$, $1.59$)\\
	&			&{\bf Wireless telecommunication services} ($0.072$, $8.88$)\\
\mr
2	&United States ($0.663$, $1.36$)		&{\bf Apparel, accessories and luxury goods} ($0.0138$, $6.55$)\\
	&United Kingdom ($0.097$, $1.70$)		&{\bf Packaged foods and meats} ($0.077$, $3.67$)\\
	&				&{\bf Apparel retail} ($0.053$, $8.28$)\\
	&				&{\bf Leisure products} ($0.039$, $3.93$)\\
	&				&{\bf Restaurants} ($0.038$, $4.53$)\\
\mr
3	&United States ($0.551$, $1.13$)		&{\bf Semiconductors} ($0.229$, $11.44$)\\
	&Taiwan ($0.084$, $8.49$)		&{\bf Technology hardware, storage and peripheral} ($0.108$, $7.94$)\\
	&Japan ($0.077$, $2.71$)			&Application software ($0.084$, $1.80$)\\
	&South Korea ($0.023$, $1.99$)		&{\bf Communications equipment} ($0.066$, $3.36$)\\
	&Israel ($0.017$, $1.70$)			&{\bf Systems software} ($0.066$, $4.78$)\\
\br
\end{tabular*}
\end{table}

\begin{table}[!ht]
\caption{\label{tabl_6}Simulated shares of conflict minerals in top $10$ industry sectors in G8 and reduction rate of conflict minerals by application of the regulation that all firms in the ``metals and mining'' upper-sector and the ``trading companies and distributors'' sector in G8 must not distribute conflict minerals to their customers; the $q_{i}$s of these firms are one and $q_{i}=0.3$ otherwise.}
\begin{indented}
\footnotesize\rm
\item[]\begin{tabular}{lrr}
\br
Industry sector in G8			&Share		&Reduction rate by policy\\
\mr
Diversified metals and mining		&$0.3907$		&--\\
Trading companies and distributors	&$0.2327$		&--\\
Gold					&$0.2251$		&--\\
Aluminum				&$0.0907$		&--\\
Commodity chemicals			&$0.0074$		&$64.9\%$\\
Electrical components and equipment	&$0.0067$		&$97.9\%$\\
Alternative carriers			&$0.0047$		&$99.1\%$\\
Silver					&$0.0039$		&--\\
Oil and gas exploration/production	&$0.0036$		&$5.6\%$\\
Diversified chemicals			&$0.0027$		&$3.7\%$\\
\mr
Other sectors				&$0.0318$		&$43.4\%$\\
\br
\end{tabular}
\end{indented}
\end{table}

\begin{table}[!ht]
\caption{\label{tabl_7}Reduction rate from amount of conflict minerals before applying regulation to $3\%$ firms in G8 to amount after. Reduction rates in G8 firms, non-applied G8 firms, firms outside G8, DRC, and DRC's neighbors are numerically estimated in each condition (see section 6).}
\begin{indented}
\footnotesize\rm
\item[]\begin{tabular}{crrr}
\br
&&&Firms outside G8, DRC,\\
&G8 firms&Non-applied G8 firms& and DRC's neighbors\\
\mr
Condition 1	&$35.0\%$&$1.3\%$			&$3.4\%$\\
Condition 2	&$43.2\%$	&$5.8\%$			&$4.0\%$\\
Condition 3	&$97.3\%$&$50.5\%$			&$12.0\%$\\
\br
\end{tabular}
\end{indented}
\end{table}

\clearpage
\section*{Figures}

\begin{figure}[!ht]
\begin{center}
\includegraphics[width=3.5in]{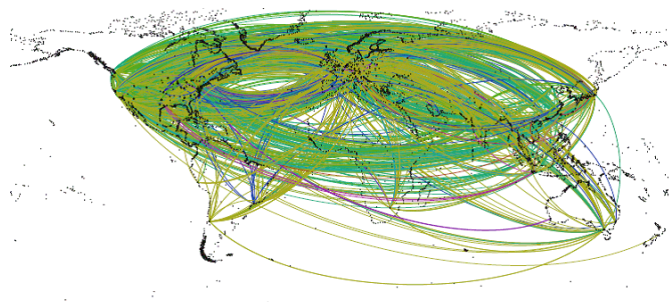}
\end{center}
\caption{\label{fig1}Global customer-supplier network in a recent period (i.e., 2013 and/or 2014). Relationships among top $1000$ firms (nodes) ranked by linkages are displayed. Each linkage color expresses a community in this network.}
\end{figure}

\begin{figure}[!ht]
\begin{tabular}{ccc}
\begin{minipage}{0.34\hsize}
\begin{center}
\includegraphics[width=2in]{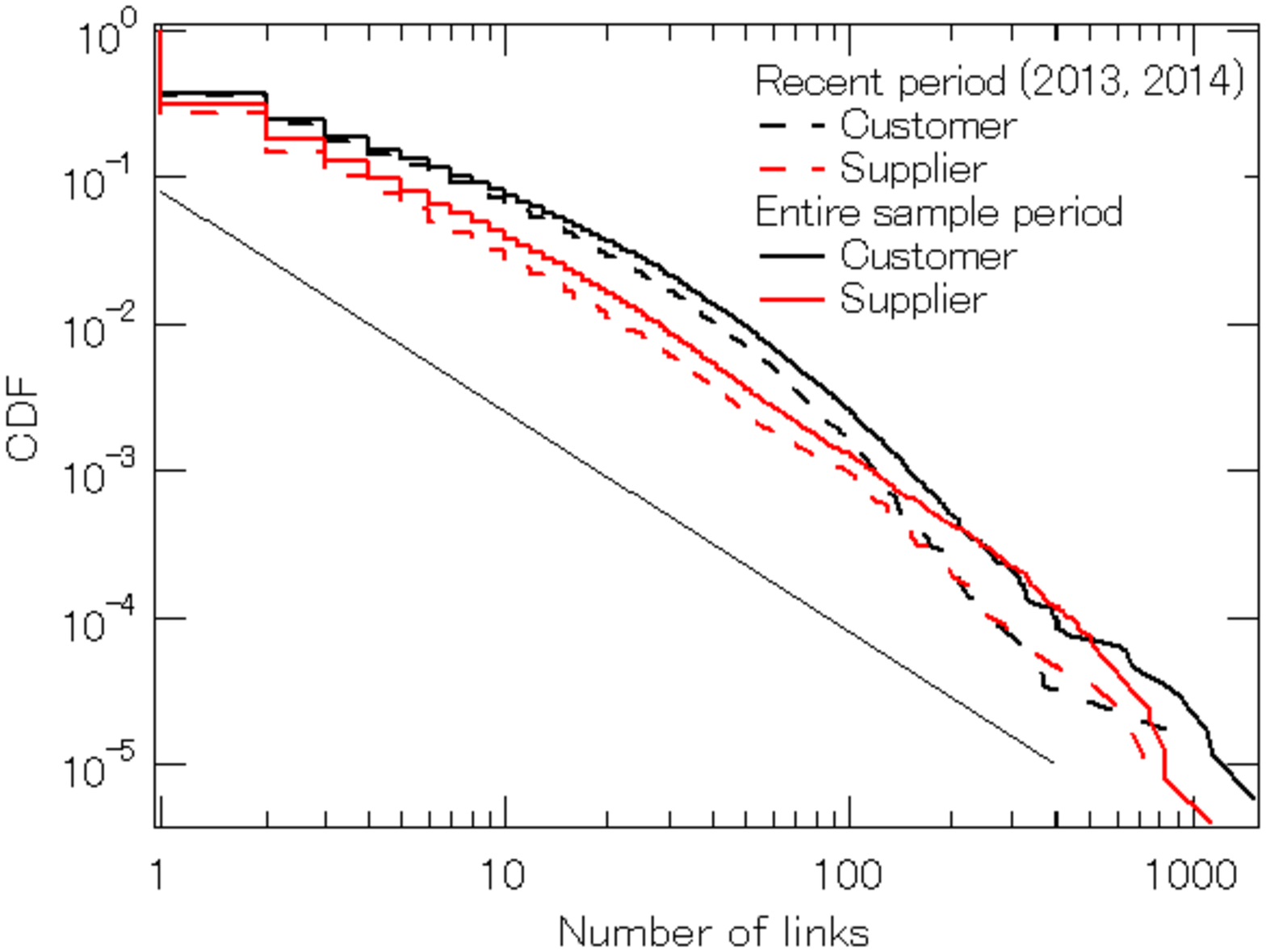}
\end{center}
\end{minipage}
\begin{minipage}{0.34\hsize}
\begin{center}
\includegraphics[width=2in]{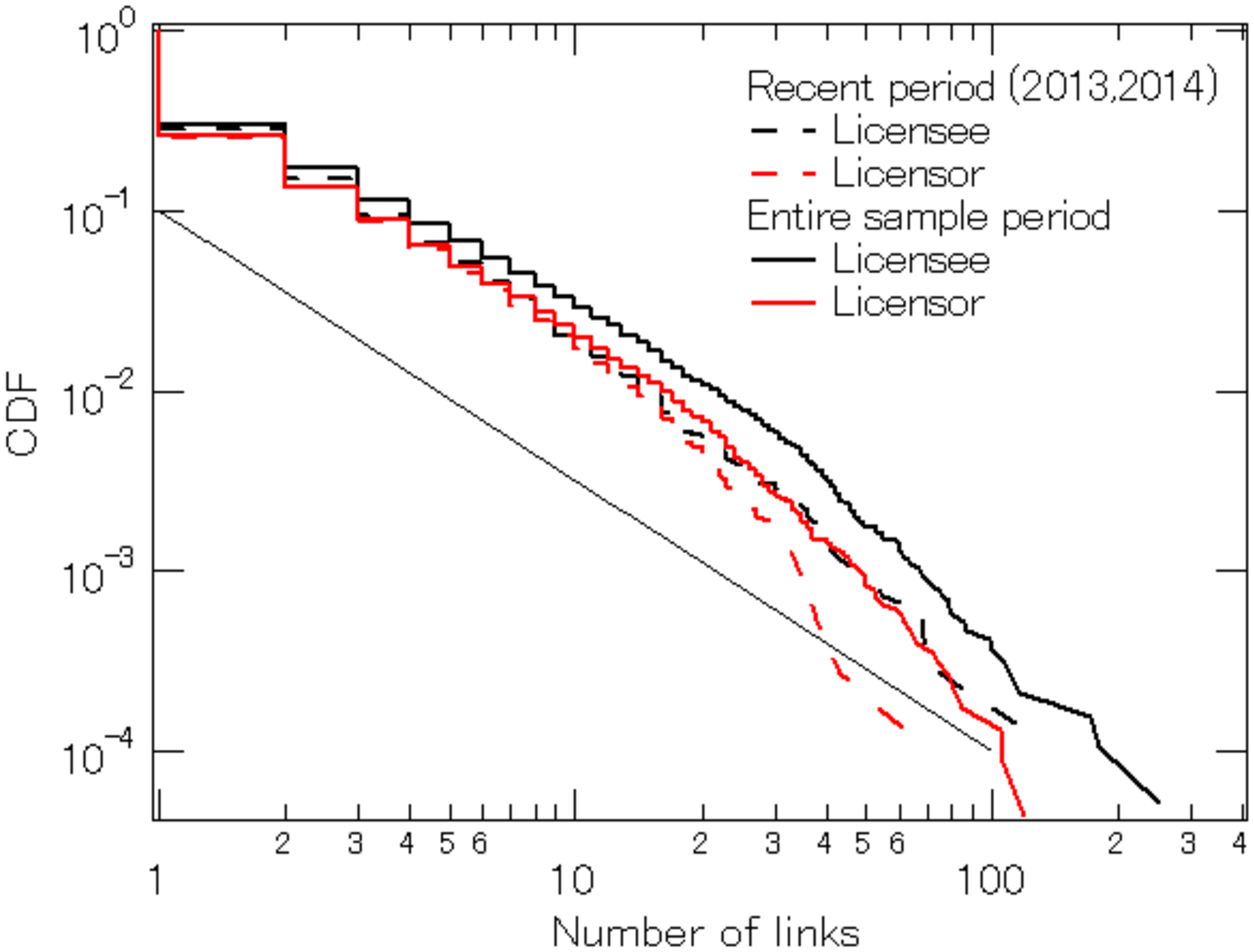}
\end{center}
\end{minipage}
\begin{minipage}{0.34\hsize}
\begin{center}
\includegraphics[width=2in]{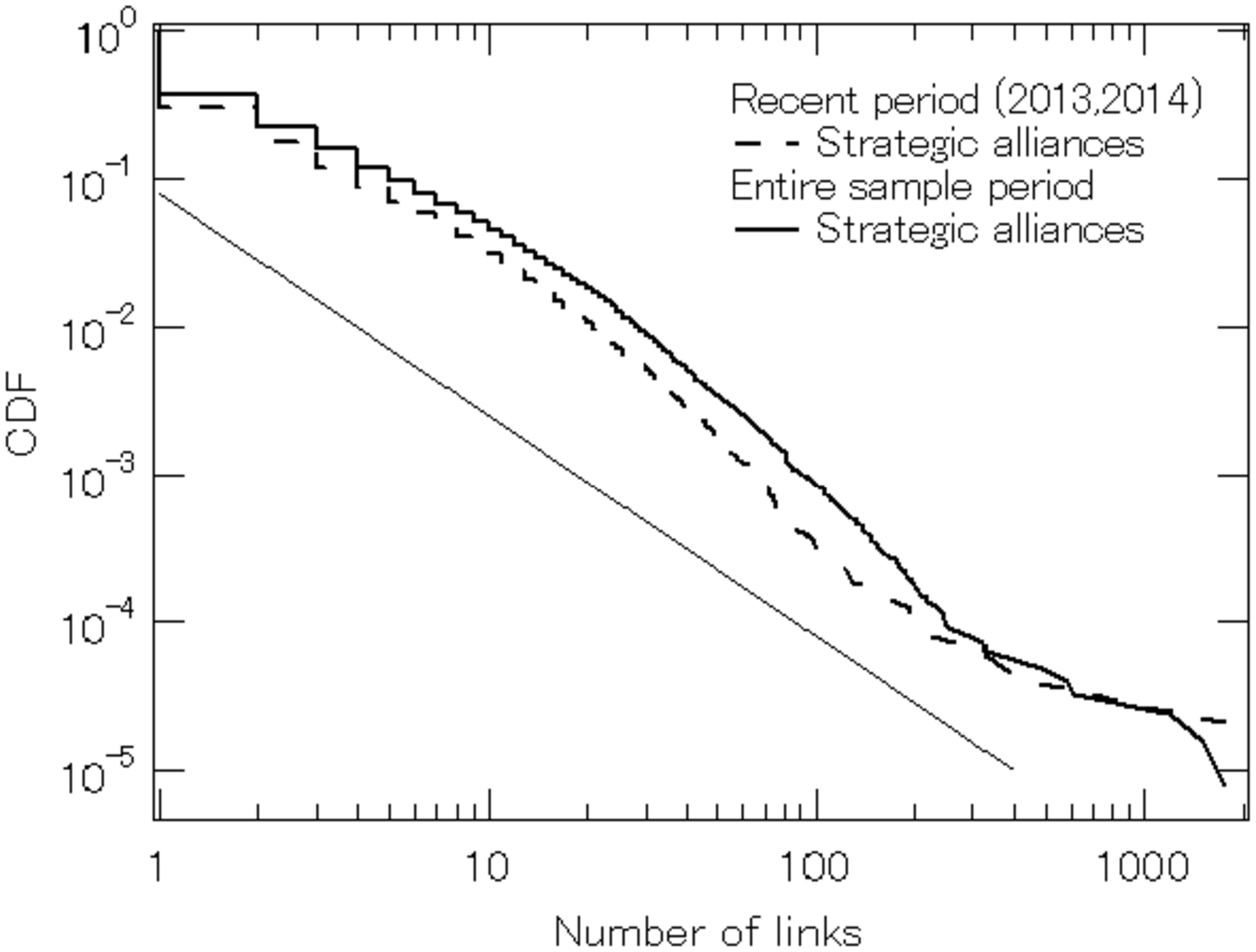}
\end{center}
\end{minipage}
\end{tabular}
\caption{\label{fig2}CDFs of links across firms for customer linkages (left, black), supplier linkages (left red), licensee linkages (center, black), licensor linkages (center, red), and strategic alliance linkages (right). Dashed lines show distribution in recent period (i.e., 2013 and/or 2014). Solid lines show distribution in some years over entire sample period. Guidelines express power law with an exponent of $1.5$.}
\end{figure}

\begin{figure}[!ht]
\begin{tabular}{cc}
\begin{minipage}{0.5\hsize}
\begin{center}
\includegraphics[width=2.5in]{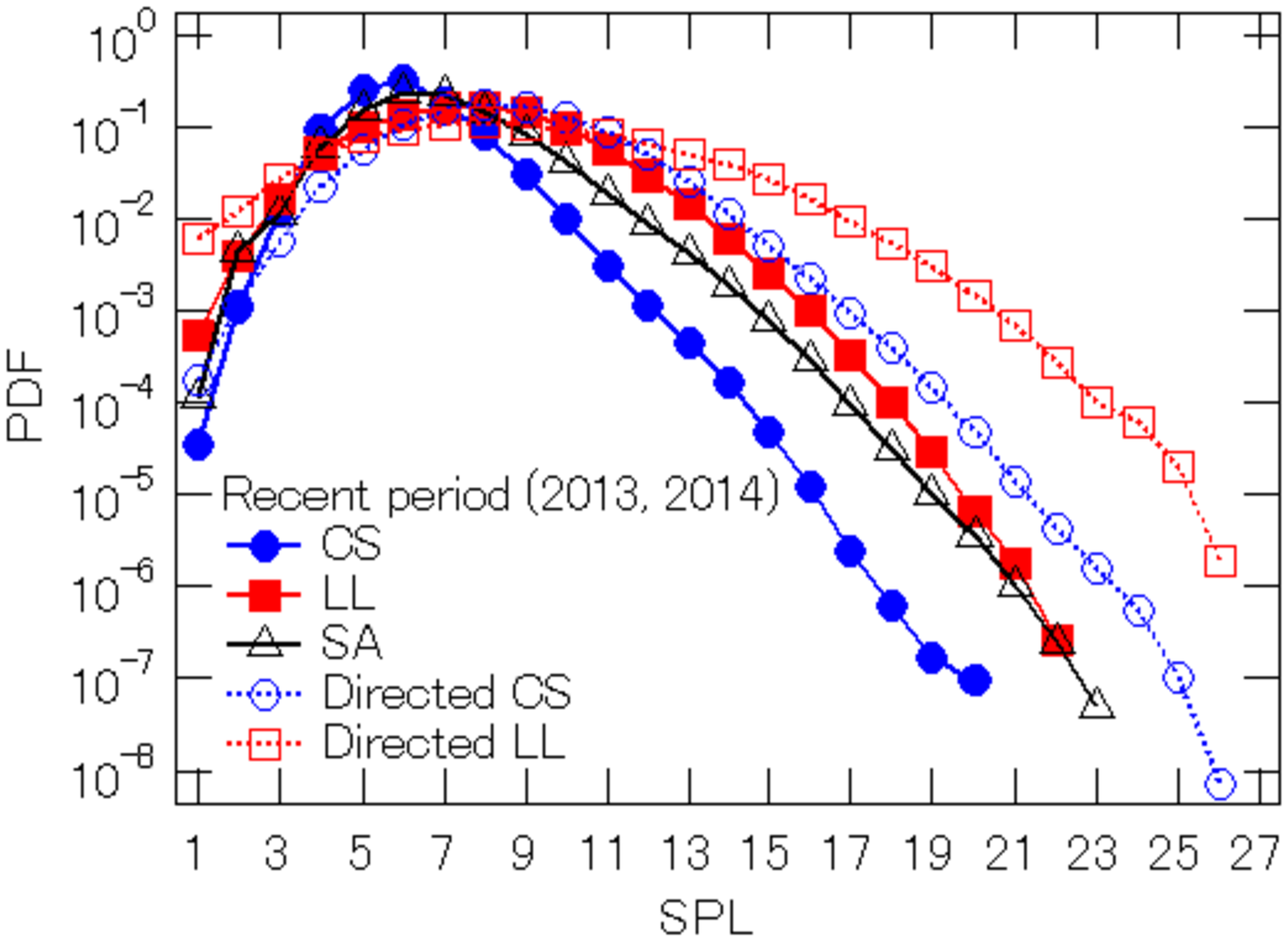}
\end{center}
\end{minipage}
\begin{minipage}{0.5\hsize}
\begin{center}
\includegraphics[width=2.5in]{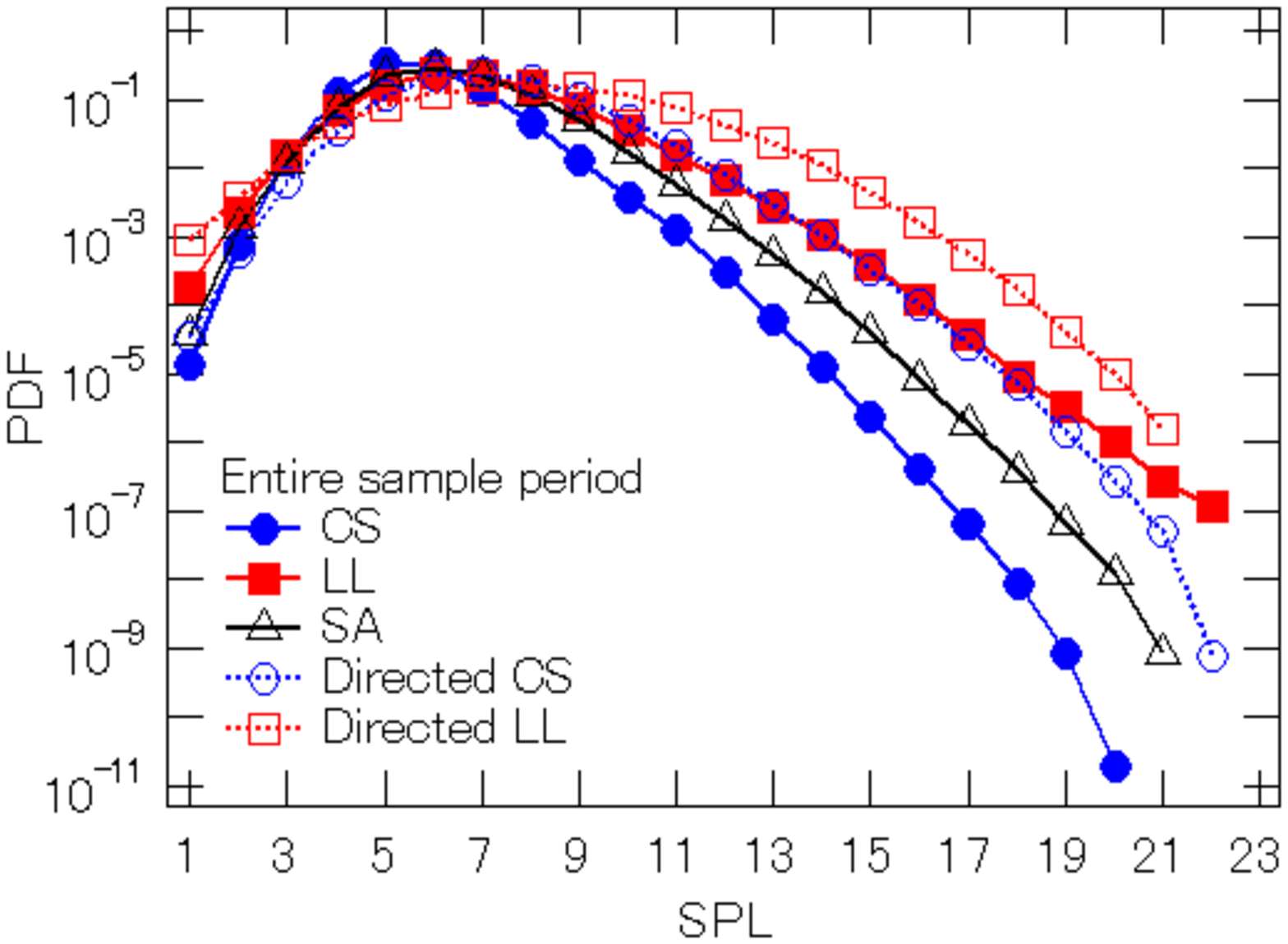}
\end{center}
\end{minipage}
\end{tabular}
\caption{\label{fig3}Distributions of shortest path lengths (SPLs) for firm pairs. Left and right figures show distributions in recent period (i.e., 2013 and/or 2014) and in some years over entire sample period. (\textcolor{blue}{\fullcircle}), (\textcolor{red}{\fullsquare}), (\opentriangle), (\textcolor{blue}{\opencircle}), and (\textcolor{red}{\opensquare}) express distributions in non-directed customer-supplier, non-directed licensee-licensor, non-directed strategic alliance, directed customer-supplier, and directed licensee-licensor networks, respectively.}
\end{figure}

\begin{figure}[!ht]
\begin{center}
\includegraphics[width=2.5in]{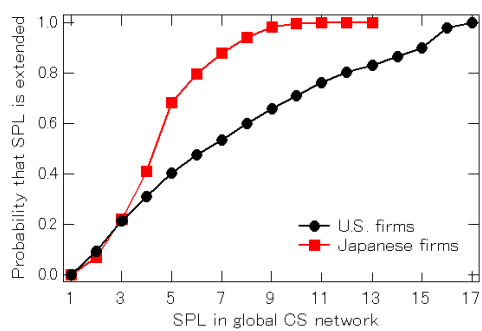}
\end{center}
\caption{\label{fig4}Probability that shortest path lengths (SPLs) are extended by being limited to domestic linkages in global customer-supplier network over entire sample period. (\fullcircle) and (\textcolor{red}{\fullsquare}) are estimation results in United States and Japan. SPL for unconnected pairs is $\infty$.}
\end{figure}

\begin{figure}[!ht]
\begin{center}
\includegraphics[width=2.5in]{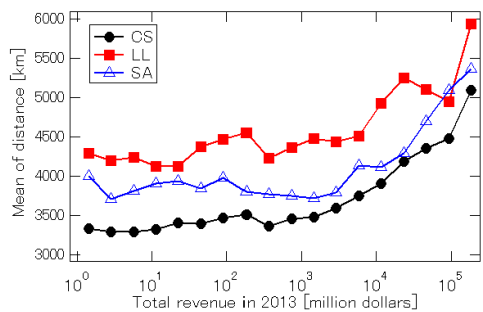}
\end{center}
\caption{\label{fig5}Relationship between firm size measured by total 2013 revenue and mean of geographical distance to business partner. (\fullcircle), (\textcolor{red}{\fullsquare}), and (\textcolor{blue}{\opentriangle}) show customer-supplier, licensee-licensor, and strategic alliance relationships in some years over entire sample period, respectively.}
\end{figure}

\begin{figure}[!ht]
\begin{center}
\includegraphics[width=4in]{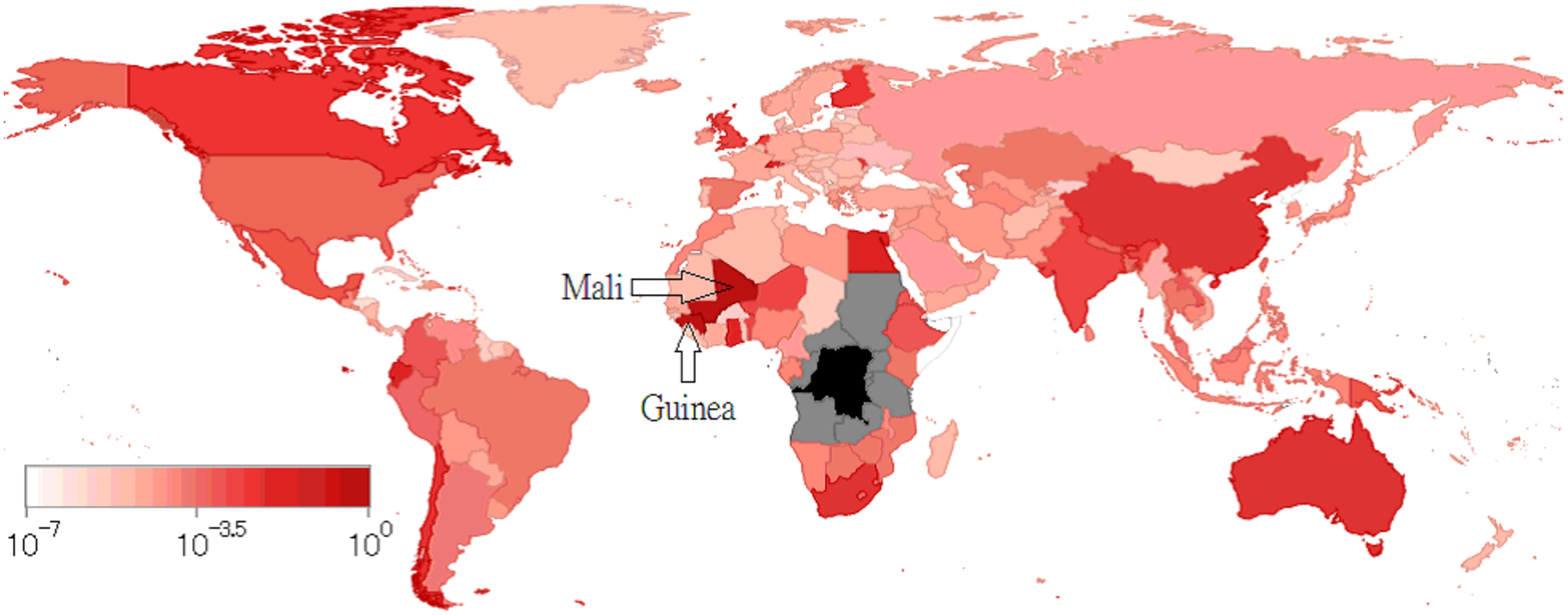}
\end{center}
\caption{\label{fig6}Simulated amount of conflict minerals per firm outside of DRC and its neighbors. Model's parameter is all $q_{i}=0.3$. Maximum amount was normalized to one. Shades of red express logarithm amount of conflict minerals. Firms in Mali and Guinea have many conflict minerals. Black expresses DRC. Gray shows its neighbors.}
\end{figure}


\section*{References}
\begin{thebibliography}{99}
\bibitem{Watts}
Watts DJ and Strogatz SH 1998 Collective Dynamics of `Small-World' Networks {\it Nature} {\bf 393} 440-442
\bibitem{Albert}
Albert R, Jeong H, and Barab\'{a}si AL 1999 Internet: Diameter of the world-wide web {\it Nature} {\bf 401} 130-131
\bibitem{Helbing}
Helbing D 2013 Globally networked risks and how to respond {\it Nature} {\bf 497} 51-59
\bibitem{Jackson}
Jackson MO 2010 {\it Social and Economic Networks} (Princeton University Press)
\bibitem{Goyal}
Goyal S 2012 {\it Connections: An Introduction to the Economics of Networks} (Princeton University Press)
\bibitem{Souma}
Souma W, Fujiwara Y, and Aoyama H 2003 Complex Networks and Economics {\it Physica \rm A} {\bf 324} 396-401
\bibitem{Saito}
Saito Y, Watanabe T, and Iwamura M 2007 Do Larger Firms Have More Interfirm Relationships? {\it Physica \rm A} {\bf 383} 158-163
\bibitem{Fujiwara}
Fujiwara Y and Aoyama H 2010 Large-Scale Structure of a Nation-Wide Production Network {\it The European Physical Journal \rm B} {\bf 77} 565-580
\bibitem{Luo}
Luo J, Baldwin CY, Whitney DE, and Magee CL 2012 The architecture of transaction networks: a comparative analysis of hierarchy in two sectors {\it Industrial \& Corporate Change} {\bf 21} 1307-1335
\bibitem{Atalay}
Atalay E, Hortacsu A, Roberts J, and Syverson C 2011 Network Structure of Production {\it Proceedings of the National Academy of Sciences} {\bf 108} 5199-5202
\bibitem{Takayasu}
Takayasu M, Sameshima S, Ohnishi T, Ikeda Y, Takayasu H, and Watanabe K 2008 Massive Economics Data Analysis by Econophysics Method-The case of companies' network structure {\it Annual Report of the Earth Simulator Center} {\bf April 2007-March 2008} 263-268
\bibitem{Glattfelder}
Glattfelder JB and Battiston S 2009 Backbone of complex networks of corporations: The flow of control {\it Phys. Rev. \rm E} {\bf 80} 036104
\bibitem{Kogut}
Kogut B and Walker G 2001 The small world of Germany and the durability of national networks {\it American Sociological Review} {\bf 66} 317-335
\bibitem{Ohnishi}
Ohnishi T, Takayasu H, and Takayasu M 2010 Network Motifs in Inter-firm Network {\it Journal of Economic Interaction and Coordination} {\bf 5} 171-180
\bibitem{Mizuno}
Mizuno T, Souma W, and Watanabe T 2014 The Structure and Evolution of Buyer-Supplier Networks. {\it PLoS ONE} {\bf 9} e100712. doi:10.1371/journal.pone.0100712
\bibitem{Newman_03}
Newman MEJ 2003 Mixing patterns in networks {\it Phys. Rev. \rm E} {\bf 67} 026126
\bibitem{Newman_04}
Newman MEJ and Girvan M 2004 Finding and evaluating community structure in networks {\it Phys. Rev. \rm E} {\bf 69} 026113
\bibitem{Iino}
Iino T, Kamehama K, Iyetomi H, Ikeda Y, Ohnishi T, Takayasu H, and Takayasu M 2010 Community structure in a large-scale transaction network and visualization {\it Journal of Physics: Conference Series} {\bf 221} 012012
\bibitem{Vitali}
Vitali S, Glattfelder JB, and Battiston S 2011 The Network of Global Corporate Control {\it PLoS ONE} {\bf 6} e25995. doi:10.1371/journal.pone.0025995
\bibitem{Bojanowski}
Bojanowski M, Corten R, and Westbrock B 2012 The structure and dynamics of the global network of inter-firm R\&D partnerships 1989-2002 {\it The Journal of Technology Transfer} {\bf 37} 967-987
\bibitem{Garlaschelli_05}
Garlaschelli D and Loffredo MI 2005 Structure and Evolution of the World Trade Network {\it Physica \rm A} {\bf 355} 138-144
\bibitem{Garlaschelli_04}
Garlaschelli D and Loffredo MI 2004 Fitness-Dependent Topological Properties of the World Trade Web {\it Phys. Rev. Lett.} {\bf 93} 188701
\bibitem{Giovanni}
Giovanni JD and Levchenko AA 2010 Putting the Parts Together: Trade, Vertical Linkages, and Business Cycle Comovement {\it American Economic Journal: Macroeconomics} {\bf 2} 95-124
\bibitem{Piccardi}
Piccardi C and Tajoli L 2012 Existence and significance of communities in the World Trade Web {\it Phys. Rev. \rm E} {\bf 85} 066119
\bibitem{He}
He J and Deem M 2010 Structure and Response in the World Trade Network {\it Phys. Rev. Lett.} {\bf 105} 198701
\bibitem{Barigozzi}
Barigozzi M, Fagiolo G, and Mangioni G 2011 Identifying the community structure of the international-trade multi-network {\it Physica \rm A} {\bf 390} 2051-2066
\bibitem{Ross}
Ross ML 2004 How do natural resources influence civil war? Evidence from thirteen cases {\it International organization} {\bf 58} 35-67
\bibitem{Billon_13}
Billon PL 2013 {\it Fuelling war: natural resources and armed conflicts} No. 373. (Routledge)
\bibitem{Billon_01}
Billon PL 2001 The political ecology of war: natural resources and armed conflicts {\it Political geography} {\bf 20} 561-584
\bibitem{Section_1502}
Section 1502, known as the ``Conflict Mineral Law'' is enforced by the Securities and Exchange Commission
\bibitem{Capital_IQ}
S\&P Capital IQ, a part of McGraw Hill Financial Inc. (http://www.spcapitaliq.com/)
\bibitem{Clauset}
Clauset A, Newman MEJ, and Moore C 2004 Finding community structure in very large networks {\it Phys. Rev. \rm E} {\bf 70} 066111
\bibitem{Watanabe}
Watanabe H, Takayasu T, and Takayasu M. 2012 Biased Diffusion on the Japanese Inter-Firm Trading Network: Estimation of Sales from the Network Structure {\it New Journal of Physics} {\bf 14} 043034. doi: 10.1088/1367-2630/14/4/043034
\bibitem{Foerster}
Foerster AT, Sarte PDG, and Watson MW 2011 Sectoral vs. Aggregate Shocks: A Structural Analysis of Industrial Production {\it Journal of Political Economy} {\bf 119} 1-38
\end{thebibliography}
\end{document}